\documentclass[12pt,a4paper]{iopart}
\pdfoutput=1

\usepackage{graphicx}

\usepackage[colorlinks=true,
	linkcolor=blue,
	urlcolor=blue,
	citecolor=blue]{hyperref}
	
\usepackage{iopams} 

\usepackage{stmaryrd}
\usepackage{dsfont}  
\usepackage{enumitem} 

\usepackage{tikz,tikz-3dplot}

\newcommand{\Sa}{%
	\begin{tikzpicture}[scale=0.15]
	\draw (0,0) rectangle (1,1);
	\end{tikzpicture}%
}%
\newcommand{\Sb}{%
	\begin{tikzpicture}[scale=0.15]
	\draw (0,0) rectangle (2,1);
	\draw[opacity=0.2] (1,1) -- (1,0);
	\end{tikzpicture}%
}%
\newcommand{\Sc}{%
	\begin{tikzpicture}[scale=0.15]
	\draw (0,0,0) -- (1,0,0) -- (1,0,-1) -- (1,1,-1) -- (1,1,0) -- (0,1,0) -- cycle;
	\draw[opacity=0.2] (1,0,0) -- (1,1,0);
	\end{tikzpicture}%
}%
\newcommand{\Sd}{%
	\begin{tikzpicture}[scale=0.15]
	\draw[opacity=0.2] (1,0,0) -- (1,1,0);
	\draw[opacity=0.2] (0,1,0) -- (1,1,0) -- (1,1,-1);
	\draw[opacity=0.2] (0,0,0) -- (0,0,-1) -- (0,1,-1);
	\draw[opacity=0.2] (0,0,-1) -- (1,0,-1);
	\draw[black] (0,0,0) -- (1,0,0) -- (1,0,-1) -- (1,1,-1) -- (0,1,-1) -- (0,1,0) -- cycle;
	\end{tikzpicture}%
}%

\newcommand{\Sh}{%
	\begin{tikzpicture}[scale=0.10]
	\draw[black] (-1,0,0) -- (1,0,0);
	\draw[black] (0,-1,0) -- (0,1,0);
        	\draw[black] (-1,1,0) -- (1,1,0) -- (1,-1,0) -- (-1,-1,0) -- cycle;
	\end{tikzpicture}%
}%

\begin{document}

\title{Squeezed vacua in loop quantum gravity}

\author{Eugenio Bianchi, 
Jonathan Guglielmon, \\
Lucas Hackl
and
Nelson Yokomizo}

\address{Institute for Gravitation and the Cosmos \& Physics Department,\\ Penn State, University Park, PA 16802, USA}

\ead{\mailto{ebianchi@gravity.psu.edu}, \mailto{jag585@psu.edu},\\ \mailto{lucas.hackl@psu.edu}, \mailto{yokomizo@gravity.psu.edu}}

\vspace{1em}

\begin{indented}
\item[] May 17,$\,$  2016  
\end{indented}

\begin{abstract}
We introduce squeezed vacua in loop quantum gravity, a new overcomplete basis of states that contain prescribable correlations between geometric operators. We study the behavior of long-range correlations  and discuss the relevance of these states for the reconstruction of a semiclassical spacetime from loop quantum gravity.
\end{abstract}

\section{Introduction}
Squeezed vacua play a central role in the description of quantum systems ranging from quantum optics \cite{Walls:1983zz} to particle production in the early universe \cite{Grishchuk:1990bj,Albrecht:1992kf}. In this article we introduce squeezed vacua for loop quantum gravity and discuss their relevance for the reconstruction of a semiclassical spacetime from the quantum theory.

In the simple case of a harmonic oscillator with vacuum $|0\rangle$ and creation operator $a^\dagger$, a coherent state is defined as a displaced vacuum $|z\rangle=e^{\,z\, a^\dagger} |0\rangle$. The complex variable $z$ represents a point in the phase space of the system and it encodes the expectation value of the position $q$ and the momentum $p$. In contrast, a squeezed vacuum of the harmonic oscillator is defined by $|\gamma\rangle=e^{\,\frac{1}{2} \gamma\, a^\dagger a^\dagger}|0\rangle$. From the parity of the state, it is clear that the expectation values of $q$ and $p$ vanish. On the other hand, the correlations $\langle q q\rangle$, $\langle p p\rangle$ and $\langle q p\rangle$ are non-trivial and are encoded in the complex variable $\gamma$. This different behavior is well-illustrated by two examples. Radio waves produced by an antenna are a coherent superposition of photons of the kind $|z\rangle$, whereas photons produced by an evaporating black hole come in pairs \cite{Hawking:1974sw} and are described by a squeezed vacuum  $|\gamma\rangle$.

Loop quantum gravity comes with a preferred vacuum, the Ashtekar-Lewandowski vacuum \cite{Ashtekar:2004eh,Rovelli:2004tv,Thiemann:2007zz}. This state describes a $3d$ quantum geometry with degenerate intrinsic metric and maximal spread in the extrinsic curvature. Most studies aimed at reconstructing a classical spacetime have focused on coherent states peaked on a point in the phase space of general relativity (or a truncation thereof) \cite{Thiemann:2000bw,Thiemann:2002vj,Bahr:2007xn,Bianchi:2009ky,Freidel:2010tt,Oriti:2011ug}. Such coherent states provide a classical background for quantum fluctuations. Uncorrelated fluctuations in the quantum geometry, however, hinder the existence of a perturbative quantum field theory phase. The difficulty is two-fold: within a definite truncation of the theory to a finite graph $\Gamma$, a general procedure for encoding correlations is needed. Moreover, in the limit of infinitely many degrees of freedom, correlations can lead to alternative representations of the holonomy-flux algebra \cite{Koslowski:2011vn,Dittrich:2014wpa}. We restrict attention to a finite-dimensional truncation of the theory\footnote{The truncation can be obtained for instance via the introduction of a cellular decomposition of the spatial manifold and the restriction to locally-flat connections on the complement of its skeleton \cite{Bianchi:2009tj,Freidel:2011ue}.} and focus on the first point. Shadow states are an attempt to address this difficulty by introducing `shadows' of Fock-like states in loop quantum gravity \cite{Varadarajan:1999it,Varadarajan:2001nm,Varadarajan:2004ui,Ashtekar:2001xp,Ashtekar:2002sn}. Graviton propagator calculations address this difficulty by matching the correlations in the boundary state to the ones in the spinfoam dynamics \cite{Rovelli:2005yj,Bianchi:2006uf,Alesci:2008ff,Bianchi:2009ri,Bianchi:2011hp,Shirazi:2015hwp}. Here we develop a general technique for addressing this problem: we introduce a set of states that span the Hilbert space of loop quantum gravity on a graph and are labeled by correlations.\\

The definition of squeezed vacua for loop quantum gravity involves three key ingredients. The first is the use of a bosonic Hilbert space as done in the spinor representation of loop quantum gravity \cite{Girelli:2005ii,Borja:2010rc,Livine:2011gp,Livine:2013wmq}. The construction is explained in Sec.~\ref{sec:bosonic}. The idea is based on Schwinger's oscillator model of spin \cite{schwinger:1952osc}: given a couple of oscillators with creation operators $a^\dagger$, $b^\dagger$ and vacuum $|0\rangle$, the state of definite spin is given by 
\begin{equation}
|j,m\rangle=\frac{(a^\dagger)^{j+m}}{\sqrt{(j+m)!}}\frac{(b^\dagger)^{j-m}}{\sqrt{(j-m)!}}|0\rangle\,.
\label{eq:}
\end{equation}
The second ingredient is the construction of an overcomplete basis of squeezed vacua for a bosonic lattice introduced in \cite{Bianchi:2015fra}, Sec.~\ref{sec:squeezed}. The third ingredient is an improvement of the original loop expansion that is at the roots of loop quantum gravity \cite{Rovelli:1989za}. Using bosonic variables, we introduce normal-ordered Wilson loops :$W_{\Sa}$: and use them to define a projector from bosonic to loop states \cite{Bianchi:2016BGHYtoapp}. We find that, as originally proposed in \cite{Ashtekar:1992tm}, loops are the quantum threads that weave a classical geometry. In Sec.~\ref{sec:expectation} we study the expectation value of local geometric operators and describe the relation between squeezed vacua, twisted geometries \cite{Freidel:2010aq,Freidel:2010bw,Bianchi:2010gc} and $U(N)$ coherent states \cite{Freidel:2010tt,Freidel:2009ck,Bonzom:2012bn}. In Sec.~\ref{sec:corr} we study correlations and show that the loop expansion provides a new perturbative tool to study long-range correlations in loop quantum gravity. We conclude the article with a discussion of perspectives on the reconstruction of a spacetime geometry from a squeezed vacuum.

\section{Bosonic representation of loop quantum gravity}\label{sec:bosonic}
Consider a graph $\Gamma$ consisting of $N$ nodes and $L$ links denoted $n$ and $\ell$ respectively. Choosing an orientation of the links of the graph, we define a set $\mathcal{S}$ consisting of $2L$ elements, two per link of $\Gamma$.  We denote the elements of $\mathcal{S}$ by an index $i=1,\ldots,2L$ and call them \emph{seeds} of the graph. Seeds are associated to end-points of links: a seed $i$ corresponds to either the source $i=s(\ell)$ or the target $i=t(\ell)$ of the oriented link $\ell$. We say that a seed belongs to a node $i\in n$ if the end-point $i$ of the link is attached to the node $n$. Two seeds belonging to the same node form a \emph{wedge} denoted $\langle i,j\rangle$.
The bosonic representation of loop quantum gravity makes use of the set of seeds $\mathcal{S}$ associated to a graph $\Gamma$. We define a bosonic Hilbert space $\mathcal{H}_{\mathcal{S}}$ with a pair of oscillators per seed. The Hilbert space of loop quantum gravity on a graph $\Gamma$ is a subspace of the bosonic Hilbert space, $\mathcal{H}_\Gamma\subset \mathcal{H}_{\mathcal{S}}$ \cite{Girelli:2005ii,Borja:2010rc,Livine:2011gp}. 

To each seed $i$ in $\mathcal{S}$ we associate a pair of bosonic degrees of freedom labeled by an index $A=0,1$. As a result, we have a bosonic system with $4L$ degrees of freedom: a \emph{bosonic lattice}. Creation and annihilation operators $a_i^A{}^\dagger, a_i^A$ satisfy the canonical commutation relations
\begin{equation}
[a_i^A,a_j^B{}^{\dagger}]=\,\delta_{ij}\,\delta^{AB},\qquad [a_i^A,a_j^B]=0,\qquad [a_i^A{}^{\dagger},a_j^B{}^{\dagger}]=0.
\label{eq:}
\end{equation}
The Hilbert space $\mathcal{H}_{\mathcal{S}}$ of the bosonic lattice is the Fock space built over the vacuum $|0\rangle$ defined as the state annihilated by all the operators $a_i^A$,
\begin{equation}
a_i^A|0\rangle=0\qquad \forall\;\,i=1,\ldots, 2L\,,\;\;A=0,1.
\label{eq:}
\end{equation}
The bosons at each seed provide a unitary representation of the group $SU(2)$ \cite{schwinger:1952osc} with generators $\vec{J}_i$ and Casimir operator $I_i$ defined at each seed $i$ by the quadratic expressions 
\begin{equation}
\vec{J}_i\equiv\frac{1}{2}\vec{\sigma}_{AB}\,a_i^A{}^\dagger\, a_i^B,\qquad I_i\equiv \frac{1}{2}\delta_{AB}\;a_i^A{}^\dagger\, a_i^B\,.
\label{eq:J&I}
\end{equation}
Here $\vec{\sigma}_{AB}$ are Pauli matrices and indices $A,B$ are always raised, lowered and contracted with the identity matrix $\delta_{AB}$. The square of the $SU(2)$ generators is  $\vec{J}_i\cdot \vec{J}_i=I_i\,(I_i+1)$. We follow the standard notation and call \emph{spins} $ j_i=0,\frac{1}{2},1,\frac{3}{2},\ldots$ the eigenvalues of $I_i$. 

To each link $\ell=\{s,t\}$ of the bosonic lattice we associate a $2\times2$ operator matrix $h_\ell$ called the \emph{holonomy} and defined as \cite{Livine:2013wmq}
\begin{equation}
h_\ell{}^A{}_B\equiv 
(2I_t+1)^{-\frac{1}{2}}\big(
\epsilon^{AC}\, a_{t\,C}^\dagger\, a_{s\,B}^\dagger
-\epsilon_{BC}\, a_t^A\, a_s^C
\big)\,
(2I_s+1)^{-\frac{1}{2}},
\label{eq:hl}
\end{equation}
where $\epsilon_{AB}$ is the $2\times 2$ antisymmetric tensor with $\epsilon_{01}=+1$. Together with the $SU(2)$ generators $\vec{J}_i$, this operator satisfies the commutation relations
\begin{equation}
[\vec{J}_s,h_\ell{}^A{}_B]=\frac{1}{2}h_\ell{}^A{}_C\,\vec{\sigma}^{C}{}_B,\qquad [\vec{J}_t,h_\ell{}^A{}_B]=-\frac{1}{2}\vec{\sigma}{}^A{}_C\, h_\ell{}^C{}_B\,.
\label{eq:}
\end{equation}
Moreover, on the subspace of $\mathcal{H}_{\mathcal{S}}$ where the condition $I_{s(\ell)}=I_{t(\ell)}$ is satisfied, the holonomy operator commutes with itself, $[(h_\ell)^A{}_B,(h_\ell)^C{}_D]=0$.
Therefore, this subspace of $\mathcal{H}_{\mathcal{S}}$ carries a representation of the holonomy-flux algebra of loop quantum gravity.

In order to reduce the bosonic Hilbert space to the loop quantum gravity Hilbert space $\mathcal{H}_\Gamma$ associated to the graph $\Gamma$, we introduce two sets of constraints: 
\begin{equation}
C_\ell\equiv I_{s(\ell)}-I_{t(\ell)}\;\approx\;0\;,\qquad \vec{G}_n\equiv \sum_{i\in n}\vec{J}_i\;\approx\;0\,.
\label{eq:link-constraint}
\end{equation}
The \emph{link constraint} $C_\ell$ imposes the matching of the spins $j_s=j_t$ at the source and target of a link $\ell=\{s,t\}$. The \emph{node constraint} $\vec{G}_n$ imposes that the coupling of the $SU(2)$ representations associated to seeds at the node $n$ is invariant under overall $SU(2)$ transformations, i.e., it is an intertwiner. These two sets of constraints can be implemented via projectors $P_\ell$ and $P_n$, so that the projector from the bosonic Hilbert space to the Hilbert space of loop quantum gravity is $P_\Gamma:\mathcal{H}_\mathcal{S}\to\mathcal{H}_\Gamma$ with $P_\Gamma=\big(\prod_{n\in \Gamma} P_n \big)\;\big(\prod_{\ell \in \Gamma} P_\ell\big)$. Via projection, any bosonic state $|s\rangle\in\mathcal{H}_\mathcal{S}$ defines a loop-quantum-gravity state
\begin{equation}
|\Gamma,s\rangle\equiv P_\Gamma|s\rangle
\label{eq:}
\end{equation} 
belonging to $\mathcal{H}_\Gamma$. In particular, the bosonic vacuum is invariant under projection $|\Gamma,0\rangle=|0\rangle$ and coincides with the \emph{Ashtekar-Lewandowski vacuum} on a graph, the eigenstate of the spin operators $I_\ell$ with vanishing eigenvalue, $I_\ell |\Gamma,0\rangle=0$. This state describes a quantum geometry with zero intrinsic metric and maximal spread in the extrinsic curvature.

\section{Squeezed vacua in loop quantum gravity}\label{sec:squeezed}
The bosonic Hilbert space $\mathcal{H}_{\mathcal{S}}$ carries a unitary representation of the symplectic group $Sp(4L,\mathbb{R})$. The generators of the symplectic algebra are
\begin{equation}
E^{AB}_{ij}\equiv \frac{1}{2}\big(a_i^A{}^\dagger\, a_j^B+a_i^A \,a_j^B{}^\dagger\big)\;,\quad F^{AB}_{ij}\equiv a_i^A\, a_j^B\;,\quad\; F^{AB}_{ij}{}^\dagger\equiv a_i^A{}^\dagger\, a_j^B{}^\dagger\,.
\label{eq:}
\end{equation}
The bosonic vacuum $|0\rangle$ is invariant up to a phase under unitary transformations generated by $E^{AB}_{ij}$. Unitary transformations generated by $F^{AB}_{ij}{}^\dagger$ \emph{squeeze} the vacuum \cite{Bianchi:2015fra}. Given a complex matrix $\gamma=(\gamma_{AB}^{ij})$ belonging to the Siegel unit-disk $\mathcal{D}$,
\begin{equation}
\mathcal{D}=\{\gamma\in \textrm{Mat}(4L,\mathbb{C})|\,\gamma=\gamma^t\;\;\textrm{and}\;\; \mathds{1}-\gamma\gamma^\dagger>0\,\}\,,
\label{eq:Siegel}
\end{equation}
we define a squeezed vacuum $|\gamma\rangle$ labeled by the squeezing parameters $\gamma_{AB}^{ij}$ and given by
\begin{equation}
|\gamma\rangle\equiv \;\exp\Big(\frac{1}{2}\, \gamma_{AB}^{ij}\, F^{AB}_{ij}{}^\dagger\Big)\,|0\rangle\,.
\label{eq:squeezed}
\end{equation}
The state has normalization $\langle\gamma|\gamma\rangle=\det(\mathds{1}-\gamma\gamma^\dagger)^{-\frac{1}{2}}$, and can be understood as a generalized coherent state for the symplectic group $Sp(4L,\mathbb{R})$ \cite{perelomov:2012}. For $\gamma=0$, we obtain the bosonic vacuum $|0\rangle$ and $\gamma\neq 0$ corresponds to a Bogoliubov transformation of the creation and annihilation operators. In fact, the squeezed vacuum $|\gamma\rangle$ is annihilated by a linear combination of $a^A_i$ and $a^A_i{}^\dagger$,
\begin{equation}
\big(a^A_i-\gamma_{AB}^{ij}\, a^B_j{}^\dagger\big)\,|\gamma\rangle=0\,.
\label{eq:}
\end{equation}
The expectation values of linear combinations of the creation and annihilation operators vanish for a squeezed vacuum: $\langle \gamma|\, a_i^A|\gamma\rangle=0$.
On the other hand, two-point correlation functions are in general non-vanishing. In particular, the spin operator $\vec{J}_i$ is a quadratic operator and  in general has non-trivial expectation value on a squeezed vacuum.


\medskip

Squeezed vacua $|\gamma\rangle$ of the bosonic lattice $\mathcal{H}_{\mathcal{S}}$ in general do not belong to the Hilbert space  $\mathcal{H}_\Gamma$ of loop quantum gravity. We define squeezed vacua in $\mathcal{H}_\Gamma$ via projection
\begin{equation}
|\Gamma,\gamma\rangle\equiv P_\Gamma |\gamma\rangle\;\in \;\mathcal{H}_\Gamma.
\label{eq:G=Pg}
\end{equation}
The projection $P_\Gamma$ can be implemented using standard techniques, for instance by expanding the state $|\gamma\rangle$ on the spin-network basis labeled by spins and intertwiners. Here we implement the projection via a loop expansion as it results in a formulation that is better suited for studying properties of squeezed vacua.

Consider an oriented loop $\alpha=\{\ell_1^{\epsilon_1},\ldots, \ell_k^{\epsilon_k}\}$ in the graph $\Gamma$, with $\epsilon_k=\pm1$. The Wilson loop operator $W_\alpha=\Tr(h_{\ell_1}^{\epsilon_1}\cdots h_{\ell_k}^{\epsilon_k})$ is defined on the bosonic Hilbert space $\mathcal{H}_{\mathcal{S}}$ via Eq.~(\ref{eq:hl}) and it commutes with the constraints (\ref{eq:link-constraint}). A multi-loop $\Phi=\{\alpha_1^{m_1},\ldots,\alpha_k^{m_k}\}$ is a collection of loops with multiplicities $m_k$. The multi-loop operator $W_\Phi=(W_{\alpha_1})^{m_1}\cdots (W_{\alpha_k})^{m_k}$ is defined as a product of Wilson loop operators. The Hilbert space of loop quantum gravity is spanned by multi-loop operators $W_\Phi$ acting on the bosonic vacuum. This is in fact the Rovelli-Smolin representation at the origin of loop quantum gravity \cite{Rovelli:1989za}. One of the difficulties of working directly in the loop representation is the existence of Mandelstam identities that relate different elements of the loop basis. Here we make use of the bosonic representation to introduce a notion of \emph{normal ordering} of Wilson loops :$W_\Phi$: and avoid this difficulty. In particular, we introduce the multi-loop operator $F_\Phi$ defined as
\begin{equation}
F_\Phi=\prod_{\alpha\in \Phi}\Big(\prod_{\langle i,j\rangle\in \alpha}\epsilon_{AB}\,a_i^A a_j^B\Big)^{m_\alpha}
\label{eq:}
\end{equation}
where $\langle i,j\rangle$ is a couple of seeds belonging to the same node in the loop $\alpha$. It is straightforward to show that :$W_\Phi$:$|0\rangle=F^\dagger_\Phi|0\rangle$. This result allows us to derive an expression for the projector $P_\Gamma$ in terms of a sum over multi-loops 
\begin{equation}
P_\Gamma=\sum_\Phi \frac{1}{\prod_\ell (2j_\ell)!\; \prod_n(j_n+1)!}\;F_\Phi^\dagger|0\rangle\langle0|F_\Phi
\label{eq:Ploop}
\end{equation}
where $2j_\ell$ is the multiplicity of the link $\ell$ in the multiloop  $\Phi$ and $j_n=\sum_{i\in n}j_i$. The sum is over the set of multi-loops that contain only non-self-repeating loops\footnote{A self-repeating loop is a loop consisting of a sequence of links such that, including the orientations, it is invariant under a cyclic permutation of length smaller than the number of links in the loop.}. The projector (\ref{eq:Ploop}) provides a loop expansion for states in $\mathcal{H}_\Gamma$ \cite{Bianchi:2016BGHYtoapp}.

To compute the loop expansion of a squeezed vacuum, we introduce a set of complex variables $z_i^A\in\mathbb{C}^2$ associated to the seeds $i\in \mathcal{S}$ of the graph $\Gamma$ and define the holomorphic function
\begin{equation}
Z_\Phi\equiv
\prod_{\alpha\in\Phi}\Big(\prod_{\langle i,j\rangle\in \alpha}\!\!\epsilon_{AB}\,z^A_i z^B_j\;\Big)^{m_\alpha}\,.
\label{eq:Zphi}
\end{equation}
The scalar product of a squeezed vacuum $|\gamma\rangle$ with a multiloop state $F_\Phi^\dagger|0\rangle$ defines a function $\mu_\Phi(\gamma)$ that can be expressed as a complex integral
\begin{equation}
\mu_\Phi(\gamma)\equiv\langle 0|F_\Phi|\gamma\rangle=
\int \bar{Z}_\Phi\;
e^{-z_i^A \bar{z}^i_A+\frac{1}{2}\gamma_{ij}^{AB}z^i_A z^j_B}\;\;{\textstyle\prod_{i,A}\frac{dz_i^A\wedge d\bar{z}_i^A}{\pi}}.
\label{eq:muPhi}
\end{equation}
Using Eq.~(\ref{eq:Ploop}) and (\ref{eq:muPhi}), we find that  squeezed vacuum states $|\Gamma,\gamma\rangle$ can be written as a superposition of multiloop states $F_\Phi^\dagger|0\rangle$ as
\begin{equation}
|\Gamma,\gamma\rangle=\sum_\Phi \frac{\mu_\Phi(\gamma)}{\prod_\ell (2j_\ell)!\; \prod_n(j_n+1)!}\;F_\Phi^\dagger|0\rangle.
\label{eq:loopexpansion}
\end{equation}
Note that squeezed vacua form an overcomplete basis of the loop quantum gravity Hilbert space $\mathcal{H}_\Gamma$. The resolution of the identity is given by
\begin{equation}
P_\Gamma=\int |\Gamma,\gamma\rangle \langle\Gamma,\gamma|\;d\mu(\gamma)
\label{eq:}
\end{equation}
where $d\mu(\gamma)$ is the Haar measure on the symplectic group $Sp(4L,\mathbb{R})$. This result follows straightforwardly from Eq.~(\ref{eq:G=Pg}) together with the expression of the resolution of the identity for bosonic squeezed vacua \cite{Bianchi:2015fra}.

The introduction of the set of states $|\Gamma,\gamma\rangle$ is the main result of this paper. In the following sections, we discuss their properties and their relevance for reconstructing a quantum space-time geometry in loop quantum gravity.

\section{Squeezed vacua with twisted geometries as expectation values}\label{sec:expectation}
A twisted geometry is a classical configuration described by a set of complex numbers
\begin{equation}
z_i^A\in\mathbb{C}^2,\qquad  i=1,\ldots, 2L,
\label{eq:}
\end{equation}
i.e., a spinor per seed $i\in \mathcal{S}$ of the graph $\Gamma$. The set of spinors can be equipped with the structure of a phase space $\mathcal{P}_{\mathcal{S}}=\times_{i=1}^{2L}\mathbb{C}^2$ by introducing the symplectic structure $\mathrm{i}\,dz_i^A\wedge d\bar{z}^A_i$. The vector $\vec{v}(z)$ associated to a spinor is defined as 
\begin{equation}
\vec{v}(z)=\frac{1}{2}\vec{\sigma}_{AB}\,\bar{z}^A\, z^B\,.
\label{eq:}
\end{equation}
Its norm is $|\vec{v}|=\sqrt{\vec{v}\cdot\vec{v}}=\frac{1}{2}\delta_{AB}\bar{z}^A z^B$. On the phase space $\mathcal{P}_{\mathcal{S}}$ we define the two sets of constraints 
\begin{equation}
\textstyle \mathcal{C}_\ell\equiv|\vec{v}(z_{t(\ell)})|-|\vec{v}(z_{s(\ell)})|\,\approx 0\;,\qquad\vec{\mathcal{G}}_n\equiv\sum_{i\in n}\vec{v}(z_i)\;\approx 0.
\label{eq:constraints}
\end{equation}
They generate respectively $U(1)$ and $SU(2)$ transformations at links and nodes of the graph, and are the classical analogue of the link and node constraints (\ref{eq:link-constraint}). The phase space $\mathcal{P}_\Gamma$ of loop quantum gravity is obtained via symplectic reduction with respect to the action of the link and node constraints, $\mathcal{P}_\Gamma=\mathcal{P}_{\mathcal{S}}/\!/(SU(2)^N\!\times\! U(1)^L)$  \cite{Freidel:2010aq,Freidel:2010bw}. 

A set of spinors $z_i^A$  satisfying the constraints (\ref{eq:constraints}) describes the intrinsic and extrinsic geometry of a collection of convex polyhedra called a twisted geometry. The construction is as follows. The set of spinors $z_i^A$ at a node determines the shape of a convex polyhedron with a fixed number of faces. The normal to a face of the polyhedron is $\vec{v}(z_i)$, its area is $|\vec{v}(z_i)|$ and the node constraint $\sum_{i\in n}\vec{v}(z_i)=0$ imposes the closure of the normals. Minkowski's theorem states that there exists a unique convex polyhedron with these properties  \cite{Bianchi:2010gc}.  The link constraint $|\vec{v}(z_{t(\ell)})|=|\vec{v}(z_{s(\ell)})|$ imposes that faces of polyhedra at the source and target of a link have matching areas. Note that this does not imply that the shape of the faces match \cite{Bianchi:2008es,Dittrich:2008ar}; the geometry is twisted. The vectors $\vec{v}(z_i)$ determine the spinor $z_i^A$ only up to a phase $e^{\mathrm{i}\,\xi_i}$. The link phase $\xi_\ell\equiv\xi_{t(\ell)}+\xi_{s(\ell)}$ encodes the extrinsic curvature between two neighboring polyhedra \cite{Freidel:2010aq}. \\


Here we show how to build a squeezed vacuum labeled by a set of spinors $z_i^A$ so that a twisted geometry is recovered as the expectation value of the geometric operators $\vec{J}_i$ and $h_\ell$. Given a set of spinors $z_i^A$ that satisfy the constraints (\ref{eq:constraints}), we define the \emph{node-wise squeezing} matrix $\gamma_0(z)$ as 
\begin{equation}
\gamma_0(z)_{ij}^{AB}=
\left\{
\begin{array}{ll}
\epsilon^{AB}\;\epsilon_{CD}\,z_i^C z_j^D \;\; & \textrm{if} \; \langle i,j\rangle\in n\,,      \\[1em]
\;0  &   \textrm{otherwise}.
\end{array}
\right.
\label{eq:gamma0}
\end{equation}
The matrix elements of $\gamma_0(z)$ are non-vanishing only if the two seeds $i,j$ belong to the same node $n$. The requirement that $\gamma_0(z)$ belongs to the Siegel unit-disk imposes the restriction  $0\leq \lambda_n<1$, where  $\lambda_n\equiv  \sum_{i\in n}|\vec{v}(z_i)|$ is a real number associated to a node of the graph $\Gamma$. 

The squeezed vacuum associated to $\gamma_0(z)$ is obtained by squeezing the Ashtekar-Lewandowski vacuum $|0\rangle$. From the node-wise structure of (\ref{eq:gamma0}) and the definition (\ref{eq:squeezed}) we find that the squeezed vacuum in the bosonic Hilbert space $\mathcal{H}_{\mathcal{S}}$ is a tensor product over nodes,
\begin{equation}
|\gamma_0(z)\rangle=\,\bigotimes_{n\in\Gamma}\Big(\sum_{j_n=0,1,2,\ldots}\!\!\sqrt{j_n+1}\;{\lambda_n}^{j_n}\;|j_n,\{\hat{z}_i\}\rangle_{{\!}_n}\;\Big),
\label{eq:node-wise}
\end{equation}
where $|j_n,\{\hat{z}_i\}\rangle_{{\!}_n}$ are $U(N)$ coherent states associated to nodes \cite{Freidel:2010tt,Freidel:2009ck,Bonzom:2012bn}  
\begin{equation}
|j_n,\{\hat{z}_i\}\rangle_{{\!}_n}\equiv\frac{1}{\sqrt{j_n!\,(j_n+1)!}}\Bigg(\,\frac{1}{2}\sum_{\langle i,j\rangle\in n}\epsilon_{AB} \hat{z}_i^A \hat{z}_j^B\;\epsilon_{CD}\,a^\dagger_i{}^C a^\dagger_{j}{}^D\Bigg)^{\!j_n}\;|0\rangle,
\label{eq:}
\end{equation}
and the spinors $\hat{z}^i\equiv z^i/\sqrt{\sum_{i\in n}|\vec{v}(z_i)|}$ are normalized so that $\sum_{i\in n}|\vec{v}(\hat{z}_i)|=1$. Properties of $|\gamma_0(z)\rangle$ follow easily from standard results about $U(N)$ coherent states. Since the states $|j_n,\{\hat{z}_i\}\rangle_{{\!}_n}$ are normalized to $1$, the norm of the squeezed vacuum is $\langle\gamma_0(z)|\gamma_0(z)\rangle=\prod_n\frac{1}{(1-\lambda_n^2)^2}$, which is consistent with the restriction $0\leq \lambda_n<1$. Geometrically, $|j_n,\{\hat{z}_i\}\rangle_{{\!}_n}$ describes the quantum state of a convex polyhedron with a fixed number of faces and an overall area $j_n$. The average shape of the polyhedron is encoded in the spinors $z_i^A$ as described above \cite{Bianchi:2010gc}. The expectation value of the total area of the polyhedron at the node $n$ is given by $\mathcal{A}_n\equiv\langle \gamma_0|\sum_{i\in n}I_i\,|\gamma_0\rangle/\langle\gamma_0|\gamma_0\rangle=\frac{2\lambda_n^2}{1-\lambda_n^2}$. Large average areas are obtained for $\lambda_n\to 1$, while the limit $\lambda_n\to 0$ corresponds to a state with the dominant contribution coming mostly from spins $j_i=0$ and $j_i=\frac{1}{2}$. The relative dispersion is $\Delta\mathcal{A}_n/\mathcal{A}_n=\frac{1}{\sqrt{2}\lambda_n}$. Note that the peakedness properties of squeezed states are very different from the ones of more familiar coherent states with a Gaussian profile in spins. The probability of finding spin $j_n$ is $p(j_n)=(1-\lambda_n^2)^2\,(j_n+1)\lambda_n^{\,2j_n}$. The profile of $p(j_n)$  is closer to the one of a thermal distribution with temperature $2\log 1/\lambda_n$. As a result, squeezed vacua do not not represent \emph{macroscopic} quantum polyhedra with classical properties, but provide a picture closer to the one of a thermal ensemble.

The state $|\gamma_0(z)\rangle\in \mathcal{H}_\mathcal{S}$ satisfies the node constraint but not the link constraint. The squeezed state $|\Gamma,\gamma_0(z)\rangle\equiv P_\Gamma |\gamma_0(z)\rangle$ in the loop quantum gravity Hilbert space can be determined using the loop projector. Using Eq.~(\ref{eq:Ploop}), we find that its expression on the loop basis is particularly simple: it is given by a superposition of multiloop states with weights that are explicitly expressed in terms of the spinors $z_i^A$ via the function $Z_\Phi$ defined in Eq.~(\ref{eq:Zphi}):
\begin{equation}
|\Gamma,\gamma_0(z)\rangle=\sum_\Phi \frac{Z_\Phi}{\prod_\ell (2j_\ell)!}\;  F^\dagger_\Phi|0\rangle.
\label{eq:}
\end{equation}
By construction, these states have the twisted geometry $z_i^A$ as expectation value.

\section{Long-range squeezing and correlations}\label{sec:corr}
The graph $\Gamma$ can be equipped with a graph-theoretical notion of distance between nodes. Here we discuss the scaling of $2$-point correlation functions with the distance and show how non-local squeezing introduces long-range correlations while preserving a twisted-geometry as expectation value.

A bosonic squeezed vacuum $|\gamma\rangle$ with generic squeezing matrix $\gamma_{ij}^{AB}$ has non-trivial correlation functions for bosonic operators given by
\begin{equation}
\hspace{-4.5em}\langle \gamma|\, a_i^A\,a_j^B|\gamma\rangle=\left(\gamma\;(\mathds{1}-\gamma^\dagger\gamma)^{-1}\right)_{ij}^{AB},
\quad \;
\langle \gamma|\, a_i^A{}^\dagger\, a_j^B|\gamma\rangle=2\left(\gamma\;(\mathds{1}-\gamma^\dagger\gamma)^{-1}\,\gamma^\dagger\right)_{ij}^{AB}.
\label{eq:<aa>}
\end{equation}
The choice of node-wise squeezing matrix (\ref{eq:gamma0}) results in non-trivial correlations only between seeds at the same node. This fact is manifest in the expression (\ref{eq:node-wise}) for the node-wise squeezed state:  the state  $|\gamma_0(z)\rangle$ is a tensor product over nodes and therefore has no correlations between the geometry at distinct nodes. The projection $|\Gamma,\gamma_0(z)\rangle=P_\Gamma|\gamma_0(z)\rangle$ introduces short-ranged correlations which fall off exponentially with the distance, but there are still no long-ranged correlations. Long-range correlations can be introduced by adding off-diagonal components to the  squeezing matrix (\ref{eq:gamma0}). For concreteness, we discuss the case of a cubic lattice.

\bigskip

Consider a cubic lattice $\Gamma$ consisting of $N$ nodes and $L=3N$ links. We assume $3$-torus topology. We label nodes by three integers $n=(n_1,n_2,n_3)$. At each node $n$ of the cubic lattice there are six seeds corresponding to the six links at the node. We denote them by an index  $\mu=\pm 1, \pm 2,\pm 3$. For instance, the seeds $\mu=+3$ and $\mu=-3$ are associated to links that connect the node with coordinates $(n_1,n_2,n_3)$ to the node with coordinates $(n_1,n_2,n_3+1)$ and $(n_1,n_2,n_3-1)$ respectively. We use the notation $i=(n,\mu)$ to identify a seed in the cubic lattice by the node $n$ and the direction $\mu$. 

We now introduce a set of spinors $z_{n\mu}^A=\sqrt{\lambda}\, \hat{z}_\mu^A$ that describe a twisted geometry consisting of a collection of identical Euclidean cubes, one per node of the graph. The intrinsic geometry is the one of a cubulation of $3d$ Euclidean space with the size of cubes fixed by the scale $\lambda$. The extrinsic curvature is determined by a set of phases $\xi_\mu$. We choose the normalized spinors $\hat{z}_\mu^A$ to be given by
\begin{equation*}
\hspace{-2em}
\begin{array}{lll}
\hat{z}_1 = e^{i \xi_1} \frac{e^{\mathrm{i}\pi/4}}{\sqrt{6}} \Bigg(\!\begin{array}{c} 1 \\ 1 \end{array}\!\Bigg),\quad
& \hat{z}_2 = e^{\mathrm{i} \xi_2} \frac{e^{-i\pi/4}}{\sqrt{6}} \Bigg(\!\begin{array}{c} 1 \\ \mathrm{i} \end{array}\!\Bigg),\quad
& \hat{z}_3 = e^{\mathrm{i} \xi_3} \frac{e^{i\pi/4}}{\sqrt{3}}  \Bigg(\!\begin{array}{c} 1 \\ 0 \end{array}\!\Bigg),\\[1.5em]
\hat{z}_{-1} =  \frac{1}{\sqrt{6}} \Bigg(\!\begin{array}{c} 1 \\ -1 \end{array}\!\Bigg),\quad
& \hat{z}_{-2} = \frac{1}{\sqrt{6}} \Bigg(\!\begin{array}{c} 1 \\ -\mathrm{i} \end{array}\!\Bigg),\quad
& \hat{z}_{-3} = \frac{1}{\sqrt{3}}  \Bigg(\!\begin{array}{c} 0 \\ 1 \end{array}\!\Bigg)
\end{array}
\end{equation*}
The vectors $\vec{v}(\hat{z}_\mu)$ are normals to the faces of a cube of unit total area,  $\sum_\mu |\vec{v}(\hat{z}_\mu)|=1$.

In order to introduce correlations between different nodes, we consider the squeezing matrix $\gamma_{(m\mu)(n\nu)}^{AB}$ defined as
\begin{equation}
\gamma_{(m\mu)(n\nu)}^{AB}=\lambda\;\big(\delta_{mn}\,+\,\varepsilon\, f_{mn}\big)\;\epsilon_{CD}\,\hat{z}_\mu^C\hat{z}_\nu^D\;\,\epsilon^{AB}\,.
\label{eq:long-range}
\end{equation}
For $\varepsilon=0$, this squeezing matrix falls in the class (\ref{eq:node-wise}) and defines a state with no long-range correlations. The term $f_{mn}$ introduces correlations between distinct nodes $m$ and $n$. The squeezed vacuum $|\Gamma,\gamma\rangle=P_\Gamma |\gamma\rangle$ is defined via the loop expansion (\ref{eq:loopexpansion}). In the small squeezing limit $\lambda \ll 1$, the loop expansion results in the series where the term of order $\lambda^k$ is a sum over loops of length $k$,
\begin{eqnarray}
\label{eq:small-squeezing}
|\Gamma,\gamma\rangle=&|0\rangle+\lambda^4\sum_{\Sa\,}c_{\Sa}\;F^\dagger_{\Sa}\,|0\rangle+\\
&\hspace{-1em}+\lambda^6\Big(\sum_{\Sb\,}c_{\Sb}\,F^\dagger_{\Sb}|0\rangle+\sum_{\Sc\,}c_{\Sc}\,F^\dagger_{\Sc}|0\rangle+\sum_{\Sd\,}c_{\Sd}\,F^\dagger_{\Sd}|0\rangle\Big)\nonumber\\
&\hspace{-1em}+\lambda^8\Big(\sum_{\Sa\,\Sa}c_{\Sa\,\Sa}\,F^\dagger_{\Sa}F^\dagger_{\Sa}|0\rangle+\ldots\Big)\;+O(\lambda^{10}).\nonumber
\end{eqnarray}
The expansion is analogous to the one studied in lattice gauge theory \cite{Drouffe:1983fv} where $\Sa$ is a \emph{plaquette} in the cubic lattice and $\Sb\,, \Sc\,, \Sd$ are loops of length $6$. The norm of the state is $\langle\Gamma,\gamma|\Gamma,\gamma\rangle = 1 + O(L\, \lambda^8)$, where $L$ is the number of links in the lattice. The state $|\Gamma,\gamma\rangle$ can thus be taken to be normalized for  $\lambda^8 \ll 1/L$.

Let us first consider the case of $\varepsilon=0$. The coefficient $c_{\Sa}$, i.e. the amplitude of a single plaquette excitation in the $\mu\,\nu$-plane, can be computed explicitly and is given by $c_{\Sa} = \frac{1}{2^2 \,3^4} \exp({\mathrm{i}\,2(\xi_\mu+\xi_\nu))}$. The expectation value of a Wilson loop for a plaquette lying in the $\mu\nu$-plane is
\begin{equation}
\langle \Gamma,\gamma|\, W_{\Sa}\, |\Gamma,\gamma\rangle  = \frac{\lambda^4}{2 \cdot 3^4} \cos(2\xi_\mu+2\xi_\nu) \,+O(\lambda^8) . 
\label{eq:mean-W}
\end{equation}
This result supports the interpretation of the phases $\xi_\mu$ in the state as parametrizing the holonomy of the Ashtekar connection. The expectation value of the spin operator $I_\ell$ on a link is 
\begin{equation}
\langle \Gamma,\gamma|\, I_\ell\, |\Gamma,\gamma\rangle = \frac{2}{3^8} \,\lambda^8 \,+O(\lambda^{12})\,.
\label{eq:mean-j}
\end{equation}
In loop quantum gravity, states with definite spin are eigenstates of the area operator. Identifying the spin operator with the area of a face of the cube, $\mathcal{A}_\ell= I_\ell$,\footnote{We use units $8\pi G \beta=1$ where $G$ is Newton's constant and $\beta$ the Barbero-Immirzi parameter. The results generalize to the operator-ordering $\mathcal{A}_\ell= \sqrt{I_\ell(I_\ell+1)}$.}  we find that the small-squeezing limit $\lambda\ll 1$ corresponds to a cubulation with cubes having  linear size $\ell_{\mathrm{cube}}(\lambda)\equiv \sqrt{\langle\mathcal{A}_\ell\rangle}=\frac{\sqrt{2}}{3^4}\,\lambda^4$. Together with the regular geometry of the $Euclidean$ cubulation, this result allows us to introduce a notion of metric distance between cubes,
\begin{equation}
 d_{nm}(\lambda)\equiv \sqrt{(n_1-m_1)^2+(n_2-m_2)^2+(n_3-m_3)^2\,}\;\,\ell_{\mathrm{cube}}(\lambda)\,.
\label{eq:}
\end{equation}
A suitable generalization of the length operator in loop quantum gravity should reproduce this distance as expectation value \cite{Bianchi:2008es,Thiemann:1996at}.

\bigskip

We now consider spins on distant links and discuss the correlation function
\begin{equation}
C_{\ell\ell'}\equiv \,\langle \Gamma,\gamma|\, I_\ell\;I_{\ell'} |\Gamma,\gamma\rangle-\langle \Gamma,\gamma|\, I_\ell\, |\Gamma,\gamma\rangle\,\langle \Gamma,\gamma|\, I_{\ell'}\, |\Gamma,\gamma\rangle.
\label{eq:}
\end{equation}
In the case $\varepsilon=0$, the correlation function scales with the distance as $C_{\ell\ell'}\sim \lambda^{k}$, where $k$ is the length of the shortest loop containing both $\ell$ and $\ell'$ (we assume $k>6$ in the following). This behavior corresponds to an exponential fall-off of correlations. On the other hand, for $\varepsilon\neq 0$, the dominant contribution to the correlation function comes from terms with two disjoint plaquettes excited, i.e., from terms $F^\dagger_{\Sa}F^\dagger_{\Sa}\,|0\rangle$. The correlation function at the leading order in $\lambda$ and $\epsilon$ is 
\begin{equation}
C_{\ell \ell'} = \frac{2^6}{3^{16}} \lambda^{16} \varepsilon^2 \,f_{s(\ell)s(\ell')}^{\;2} \, 
\label{eq:}
\end{equation}
for two parallel links $\ell$, $\ell'$ in the cubic lattice. In particular, by choosing the function $f_{mn}$ in Eq.~(\ref{eq:long-range}) to scale with the inverse distance, $f_{mn}\propto 1/d_{mn}(\lambda)$, we find that spins on distant links are correlated and the correlation function scales as 
\begin{equation}
C_{\ell \ell'} \propto 1/(d_{s(\ell)s(\ell')})^2.
\label{eq:1/d2}
\end{equation}
The inverse-distance-squared scaling of the correlation function in a squeezed vacuum reproduces the characteristic decay of equal-time correlations of vacuum fluctuations for a massless quantum field in Minkowski space. Note that the distance between the links $\ell$, $\ell'$ is not measured with respect to a background geometry but it is encoded in the state itself.

\section{Discussion: reconstructing a spacetime geometry}
Loop quantum gravity is often presented in terms of spin-network states, eigenstates of the intrinsic quantum geometry of space. The squeezed vacua $|\Gamma,\gamma\rangle$ introduced in this article, Eq.~(\ref{eq:loopexpansion}), provide a new overcomplete basis of loop quantum gravity that is tailored to the study of a classical spacetime geometry and its quantum fluctuations.

Most studies of semiclassicality in loop quantum gravity are based on coherent states labeled by a point in the phase space of the theory. Squeezed vacua also reproduce classical configurations as expectation values, but they do more than that: they encode long-range correlations that ordinary coherent states cannot capture. Squeezed states are labeled by correlation functions of $z_i^A$ with $z_j^B$ in a twisted geometry. Correlations at the same node or nearby nodes encode the expectation value of local geometric operators as the spin or the holonomy and provide a classical background. Correlations between distant nodes encode quantum fluctuations over that background.

Squeezed vacua $|\Gamma,\gamma\rangle$ are kinematical states, they are not required to solve the Wheeler-deWitt equation. They provide, however, a new tool for finding approximate semiclassical solutions. Let us consider a definite proposal for the Hamiltonian constraint operator $H$ in loop quantum gravity. In a cubic non-graph-changing setting for instance \cite{Thiemann:2007zz}, the Hamiltonian constraint is a sum over nodes and it acts locally at each node by constraining the degrees of freedom on all the $12$ plaquettes containing the node (represented here as $\Sh$).\footnote{Models with an ultra-local action at nodes are not considered here as they lead to no propagating degrees of freedom. See \cite{Smolin:1996fz}.} Diagrammatically, we have $H=\sum_n N_n H_{\Sh}$, where $N_n$ is a lapse function and physical states satisfy $H_{\Sh}|\textrm{phys}\rangle=0$. While a squeezed state $| \Gamma,\gamma\rangle$ does not solve the Hamiltonian constraint, its parameters can be chosen so that the expectation value of $H_{\Sh}$ vanishes. This logic is often used for coherent states: the classical parameters of a coherent state are chosen so that they select a configuration in the truncated phase space of general relativity that satisfies the classical Hamiltonian constraint. Using squeezed states, we can push this strategy one step forward and require that both the expectation value and the correlation function of the Hamiltonian constraint vanish on a squeezed vacuum
\begin{equation}
\langle \Gamma,\gamma| H_{\Sh} | \Gamma,\gamma\rangle\simeq 0\,,\qquad \langle \Gamma,\gamma| H_{\Sh}\;H_{\Sh\,{}'} | \Gamma,\gamma\rangle\simeq 0\qquad \forall \,\;\Sh\,,\Sh\,{}'.
\label{eq:}
\end{equation}
This strategy selects a subset of squeezing matrices $\gamma$ that identify approximate physical states. The node-wise components $\gamma_{\langle i,j\rangle}^{AB}$ identify a classical geometry providing a background configuration, the components $\gamma_{ij}^{AB}$ associated to distinct nodes restrict the quantum fluctuations to be physical. This procedure can be understood as the first two orders in a perturbative scheme where all the $n$-point correlation functions of $H_{\Sh}$ are imposed to vanish. The requirement  $\langle\Gamma,\gamma| H_{\Sh}\;H_{\Sh\,{}'} | \Gamma,\gamma\rangle\simeq 0$ is crucial for obtaining a regime effectively described by a perturbative quantum field theory with long-ranged correlations on a classical background. The pertubative analysis presented in Sec.~\ref{sec:corr} shows that states (\ref{eq:small-squeezing}) for a cubic lattice can encode a Minkowski background geometry with quantum fluctuations having correlations that fall off as the inverse-distance-squared as expected for gravitons in the Minkowski vacuum state.

In the covariant approach to loop quantum gravity, states are associated to the boundary of a spacetime region \cite{Rovelli:2004tv,Oeckl:2005bv} and the dynamics is encoded in the spinfoam path-integral \cite{Engle:2007wy,Freidel:2007py}. A semiclassical spacetime is reconstructed via the choice of a boundary state with prescribed one-point functions for local boundary observables. Two-point correlation functions of boundary observables probe the propagation of quantum fluctuations in the bulk. Already at the level of a single vertex it is manifest that a non-trivial graviton propagator arises only if, besides peakedness, the boundary state encodes correlations between distinct nodes. The states used in \cite{Rovelli:2005yj,Bianchi:2006uf,Alesci:2008ff,Bianchi:2009ri,Bianchi:2011hp,Shirazi:2015hwp} can be considered as an early rudimentary version of the squeezed vacua introduced in this article.

The relation between Regge's triangulated spacetimes and spinfoams provides a useful guide in the study of the classical limit of the theory \cite{Regge:1961px,Barrett:2009mw}. A classical twisted geometry $z_i^A$ reproduces a triangulation only if shape-matching conditions are imposed. At the level of expectation values, the shape-matching conditions can be easily satisfied by choosing a state peaked on a triangulation \cite{Freidel:2010aq,Bianchi:2008es,Dittrich:2008ar}. In general, however, quantum fluctuations are uncorrelated and therefore do not respect this condition. By choosing appropriately the squeezing matrix $\gamma_{ij}^{AB}$, squeezed vacua can have shape-matched fluctuations. This condition is generally expected to be relevant for the existence of propagating degrees of freedom.

The bosonic representation of loop quantum gravity uses local degrees of freedom associated to seeds. Thanks to this structure, the Hilbert space $\mathcal{H}_{\mathcal{S}}$ naturally decomposes into a tensor product over regions $\mathcal{R}=\{n_1,\ldots,n_k\}$ and it allows for an unambiguous definition of the entanglement entropy $S_\mathcal{R}[|s\rangle]$ of a state restricted to a region \cite{Casini:2013rba}. The Ashtekar-Lewandowski vacuum has vanishing entanglement entropy $S_\mathcal{R}[|\Gamma,0\rangle]=0$, as is the case for all states of the spin-network basis $S_\mathcal{R}[|\Gamma,j_\ell,\mathrm{i}_n\rangle]=0$. The reason for this ultra-local behavior is that these are eigenstates of ultra-local operators that measure the quantum geometry of space at a node. On the other hand, squeezed vacua are in general non-local as the loop expansion (\ref{eq:loopexpansion}) shows. Squeezed vacua with long-range correlations as in Eq.~(\ref{eq:1/d2}) have entanglement entropy that scales as the area of the boundary of the region.\footnote{The area-law behavior is present even if we smear out the interface between the region $\mathcal{R}$ and its complement by using the mutual information.} This is the characteristic behavior expected for quantum fields on a classical background \cite{Sorkin:1983a,Srednicki:1993im}. An area-law scaling of the entanglement entropy is conjectured to be a probe of semiclassicality of states in quantum gravity \cite{Bianchi:2012ev} and the squeezed vacua introduced in this article provide a concrete illustration of this phenomenon.

\ack
We thank Abhay Ashtekar, Wolfgang Wieland and Bekir Bayta\c{s} for numerous discussions on coherent and squeezed states. The work of EB is supported by the NSF grant PHY-1404204. NY acknowledges support from CNPq, Brazil.

\section*{References}


\end{document}